\documentclass{pasj00}

\Received{$\langle$reception date$\rangle$}
\Accepted{$\langle$acception date$\rangle$}
\Published{$\langle$publication date$\rangle$}
\SetRunningHead{Y. Soejima et al.}{Photometric Studies of a WZ Sge-type dwarf nova candidate, ASAS 160048-4846.2.}

\begin{document}

\title{Photometric Studies of a WZ Sge-Type Dwarf Nova Candidate,ASAS160048-4846.2}
\author{
Yuichi \textsc{Soejima}$^1$, 
Akira \textsc{Imada}$^2$,
Daisaku \textsc{Nogami}$^3$,
Taichi \textsc{Kato}$^1$ and
L. A. G. Berto \textsc{Monard}$^4$
}

\affil{$^1$Department of Astronomy, Faculty of Science, Kyoto
University, Sakyo-ku, Kyoto 606-8501} 

\email{soejima@kusastro.kyoto-u.ac.jp}

\affil{$^2$Space Research Course in the Department of Physics, Faculty of Science, Kagoshima
University, Korimoto, Kagoshima 890-0065} 

\affil{$^3$Kwasan Observatory, Kyoto University, Yamashina-ku, Kyoto 607-8471} 

\affil{$^4$Bronberg Observatory, PO Box 11426, Tiegerpoort 0056, South Africa}


\maketitle

\begin{abstract}

We report on our time-resolved CCD photometry during the 2005 June superoutburst
 of a WZ Sge-type dwarf nova candidate, ASAS 160048-4846.2.
The ordinary superhumps underwent a complex evolution during the superoutburst.
The superhump amplitude experienced a regrowth, and had two peaks. 
The superhump period decreased when the superhump amplitude reached to
the first maximum,  successively gradually increased until the
second maximum of the amplitude, and finally decreased again. 
Investigating other SU UMa-type dwarf novae which show an increase of
the superhump period, we found the same trend of the superhump evolution
in superoutbursts of them.
We speculate that the superhump regrowth in the amplitude has a close
relation to the increase of the superhump period, and all of SU UMa-type
dwarf novae with a superhump regrowth follow the same evolution of the
 ordinary  superhumps as that of ASAS 160048-4846.2.

\end{abstract}

\section{INTRODUCTION}
\label{intro}

Cataclysmic variables (CVs) are close binaries containing a white dwarf (primary) 
and a late-type star (secondary). The secondary fills its Roche-lobe and 
transfers gas to the primary, so that an accretion disk is formed around it 
(for a review, see e.g. \cite{war95book}; \cite{hel01book}; \cite{smi07review}). 
Dwarf novae are a subclass of CVs.

SU UMa-type stars are a
subgroup of dwarf novae which show two types of outbursts: normal
outbursts, and superoutbursts. During the superoutbursts, repetitive
modulations with an amplitude of 0.1-0.3 mag, called 
superhumps, are shown. The period of the superhumps is a few percent longer than the
orbital period of the system.
The thermal-tidal instability model is the most acceptable one 
for explaining the general behavior of SU UMa-type dwarf novae
\citep{osa89suuma}. According to the tidal-instability theory, an
accretion disk becomes unstable due to the gravitational interaction
with the secondary star when it reaches a critical radius of the 3:1 resonance \citep{whi88tidal}. Superhumps can be 
explained by a beat phenomenon of the precession of a tidally-distorted
disk and the orbital motion.

WZ Sge-type dwarf novae are one of the subtypes 
of SU UMa-type dwarf novae with the shortest orbital periods among 
SU UMa stars (see e.g. \cite{kat01hvvir}). Their observational properties
are (1) the extremely long supercycle (over 5 years),
which is a period between two successive superoutbursts,
(2) the large amplitude of superoutbursts over 
6 mag (4-5 mag in many SU UMa-type stars), (3) the absence of normal
outbursts, and (4) the presence of double-peaked humps,
called early superhumps, before emergence of the ordinary superhumps.
Although the clear definition of the WZ Sge type is still controversial, its representative members where at least
  two outbursts have been observed so far and the supercycle was
  measured are AL Com (\cite{kat96alcom},
\cite{pat96alcom}), EG Cnc (\cite{kat04egcnc}), HV Vir (\cite{ish03hvvir}), WZ Sge itself
(\cite{kat04vsnet}), and GW Lib (Imada et al. in preparation).

On 2005 June 9, ASAS 160048-4846.2 (hereafter ASAS 1600) was initially 
discovered by the All Sky Automated Survey (\cite{poj02asas3}) as an eruptive
object. This is the only one recorded superoutburst. In the light
 curves of this superoutburst, \citet{ima06asas1600letter}, hereafter called Paper I, found double-peaked humps with a period of 0.063381(41) days,
 before ordinary superhumps emerged with a period of 0.064927(3)
 days. Based on evidence for early superhumps, they identified ASAS
 1600 as a promising candidate for WZ Sge-type dwarf novae.

In this paper, we reanalysed the data reported in Paper I, and
report on the evolution of the ordinary superhumps. The details
of the observations are described in the next section. The
results of the observations and the analyses are summarized in
the section 3. We will discuss the superhump evolution and
the WZ Sge nature of ASAS 1600 in the section 4, and put
conclusions in the last section 5.

\section{OBSERVATION}

Time-resolved CCD photometry was carried out in 12 consecutive
nights between 2005 June 9 and June 20 at Tiegerpoort (South Africa) 
using a 32 cm telescope.
The exposure time was 30 s with a readout-time of a few seconds. 
We used no filter during our run, so that the resultant data are close
to those of the $R_{\rm c}$ system.
After excluding bad data, we used 10511 datapoints for the following analyses. 
The journal of the observations is summarized in table \ref{log}.

After dark-subtraction and flat-fielding, we performed aperture
photometry using AIP4WIN, which is an image-editing software. 
Differential photometry was carried out using UCAC2 160053.1-484433 ($R$=11.9) as a comparison star, 
whose constancy was checked by UCAC2 160043.6-484628($R$=12.8). 
 
The heliocentric correction to the observation times was applied before the following analyses.

\begin{table*}
\begin{center}
\caption{Log of observation.}
\label{log}
\begin{tabular}{lccc}  
\hline \hline
  Date & HJD-2400000(start) & HJD-2400000(end) & $N^*$  \\ 
\hline  
  2005 June 9 & 53531.3679 & 53531.6275 & 587 \\ 
  2005 June 10 & 53532.3213 & 53532.4210 & 274 \\
  2005 June 11 & 53533.3779 & 53533.6062 & 648 \\
  2005 June 12 & 53534.2413 & 53534.6269 & 1087 \\
  2005 June 13 & 53535.2450 & 53535.5638 & 900 \\
  2005 June 14 & 53536.3708 & 53536.5904 & 565 \\
  2005 June 15 & 53537.2351 & 53537.5609 & 920 \\
  2005 June 16 & 53538.1772 & 53538.5896 & 1151 \\
  2005 June 17 & 53539.1764 & 53539.5496 & 1032 \\
  2005 June 18 & 53540.1777 & 53540.5589 & 1061 \\
  2005 June 19 & 53541.1757 & 53541.5859 & 1150 \\
  2005 June 20 & 53542.1793 & 53542.5915 & 1136 \\ 
\hline

\multicolumn{4}{@{}l@{}}{\hbox to 0pt{\parbox{180mm}{\footnotesize
\par\noindent
\footnotemark[$*$] Number of frames.
}\hss}}

\end{tabular}
\end{center}
\end{table*}

\section{RESULTS}
\label{sh}

\begin{figure}
\begin{center}
\FigureFile(80mm,!){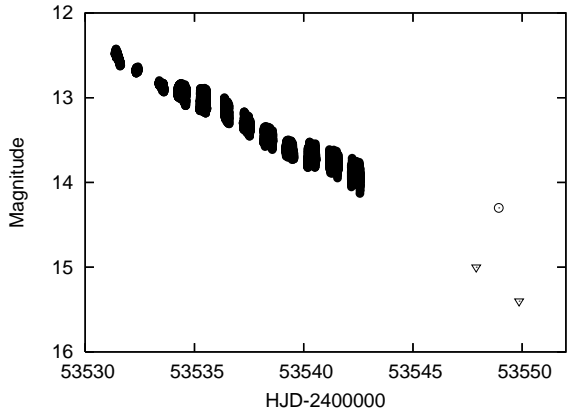}
\end{center}
\caption{Light curve of the 2005 June outburst. The abscissa is HJD,
 and the ordinate is the R band magnitude(filled circles), or the visual
 magnitude (open circles and open triangles). The filled circles
 represent our data. The open circles and the open triangles are
 the visual observations and the upper limits respectively.}
\label{lightcurve}
\end{figure}

The resulting light curve of the 2005 superoutburst of ASAS 1600 is
presented in figure \ref{lightcurve}. The object had a plateau stage, 
fading at a constant rate of 0.12 mag day$^{-1}$, which lasted at least
12 days. 
After the end of the plateau stage, ASAS 1600 underwent a rebrightening 
outburst with the maximum magnitude $V$=14.3, which was observed on 
HJD 2453548 by R. Stubbings (vsnet-outburst 6503). However, we have no
other information about the rebrightening.

\begin{figure}
\begin{center}
\FigureFile(80mm,!){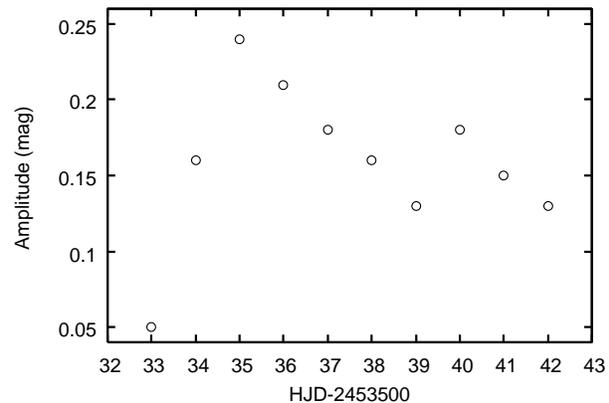}
\end{center}
\caption{Diagram of daily averaged superhump amplitude estimated by eye.}
\label{shamp}
\end{figure}

\begin{figure}
\begin{center}
\FigureFile(80mm,!){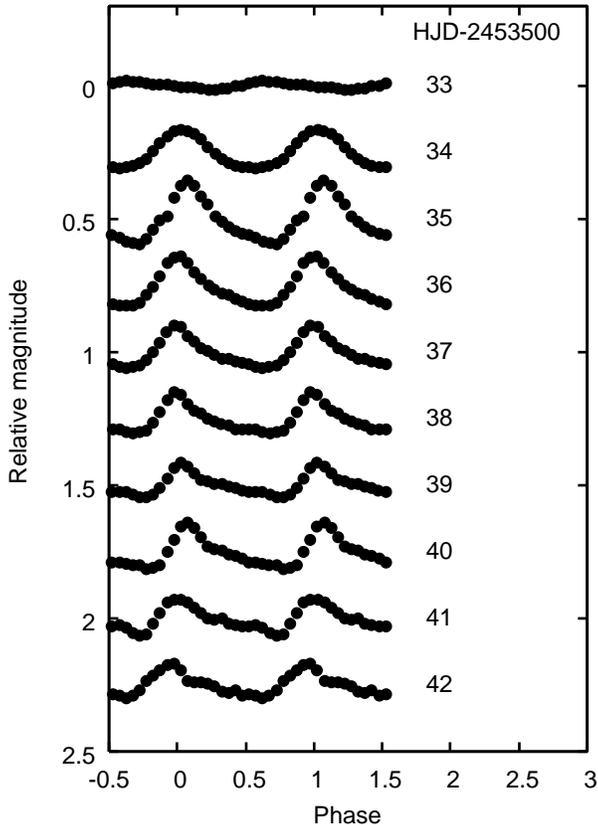}
\end{center}
\caption{Daily phase-averaged light curves of the ordinary superhump stage, folded by 0.064927 days.}
\label{dailylc}
\end{figure}

During the 2005 superoutburst, clear ordinary superhumps were detected between HJD 2453534 and HJD 2453542. On HJD 2453533, we can see humps which hint
 growth into the ordinary superhumps (see figure 4 of Paper I). 
Figure \ref{shamp} represents the change of the amplitude of the
ordinary superhumps during the 2005 superoutburst. We estimated the
amplitude by eye, and its typical error is an order of 0.01 mag.
The superhump amplitude reached to the first maximum on HJD 2453535, and
 gradually declined until HJD 2453539. However, it regrew and
reached to the second maximum on HJD 2453540. 
Afterword, the amplitude decreased. That indicates the
regrowth of the superhump amplitude occurred.
Figure \ref{dailylc} shows the daily phase-averaged light curves folded by
0.064927(3) days, which is the period of the ordinary superhumps
($P_{SH}$) measured in Paper I.

We measured the 
maximum times of the superhumps by eye (table \ref{maxtime}). The cycle
count ($E$) was set to be 1 at the first observed superhump maximum. A
linear regression to the observed maximum timings is 
represented by the following equation:

\begin{equation}
\label{c}
HJD_{max}=0.06496(2) \cdot E+2453533.4517(14).
\end{equation} 

\begin{figure}
\begin{center}
\FigureFile(80mm,!){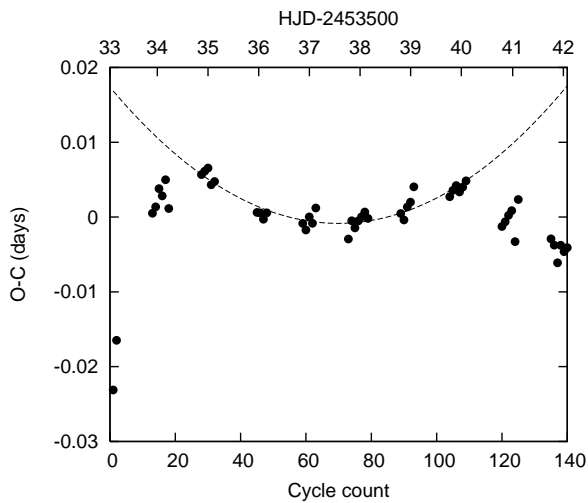}
\end{center}
\caption{$O-C$ diagram of the superhump maximum timings listed in
 table \ref{maxtime}. The curved line is obtained by a quadratic polynomial
 fitting to $O-C$ (equation \ref{o-c1}).}
\label{o-c}
\end{figure}

Using this equation, we drew an $O-C$ diagram for the maximum timings of
the superhumps in figure \ref{o-c}. The typical error is an 
order of 0.002 days, which will not affect the main results. 
As can be seen in this figure, the behavior of the $P_{SH}$ consists of
three phases. At the first and the last stage of our run, corresponding 
to around $E=30$ and $E=100$, the superhumps period shows a decreasing
trend. However, in the middle stage (about $30<E<100$), the superhump 
period clearly increases. We fitted the $O-C$ diagram
 between $28<E<109$ by the following quadratic,

\begin{eqnarray}
\label{o-c1}
O-C &=& 3.74(31) \times 10^{-6} \cdot E^{2} - 5.22(43) \times 10^{-4} \cdot E \nonumber \\
     && +1.74(14) \times 10^{-2}.
\end{eqnarray}

From this equation, the mean change rate of the superhump period between
$28<E<109$ is estimated to be 
$P_{dot}=\dot{P}_{SH}/P_{SH}=11.5(9) \times 10^{-5}$ days cycle$^{-1}$.  
Such evolution of the superhump period can be seen in some SU UMa-type
dwarf novae with short orbital period (e.g. HV Vir, \cite{ish03hvvir}).

\begin{table*}
\begin{center}
\caption{Timings of the superhump maxima.}
\label{maxtime}
\begin{tabular}{cccccc}  
\hline \hline
 $E^{*}$ & HJD-2400000 & $O-C^{\dagger}$ (days) & $E^{*}$ & HJD-2400000 & $O-C^{\dagger}$ (days) \\ 
\hline 

1 & 53533.4935 & -0.0231    & 77 & 53538.4533 & 0.0000 \\
2 & 53533.5651 & -0.0165    & 78 & 53538.5189 & 0.0007    \\  
13 & 53534.2966 & 0.0005    & 79 & 53538.5830 & -0.0002  \\   
14 & 53534.3624 & 0.0014    & 89 & 53539.2332 & 0.0005   \\   
15 & 53534.4298 & 0.0038    & 90 & 53539.2973 & -0.0004  \\   
16 & 53534.4938 & 0.0028    & 91 & 53539.3640 & 0.0014   \\   
17 & 53534.5609 & 0.0050    & 92 & 53539.4296 & 0.0020   \\   
18 & 53534.6220 & 0.0011    & 93 & 53539.4966 & 0.0040   \\   
28 & 53535.2761 & 0.0057    & 104 & 53540.2098 & 0.0027  \\   
29 & 53535.3415 & 0.0061    & 105 & 53540.2756 & 0.0036  \\   
30 & 53535.4069 & 0.0066    & 106 & 53540.3412 & 0.0042  \\   
31 & 53535.4696 & 0.0043    & 107 & 53540.4053 & 0.0034  \\    
32 & 53535.5350 & 0.0047    & 108 & 53540.4709 & 0.0040  \\    
45 & 53536.3753 & 0.0006    & 109 & 53540.5367 & 0.0048   \\   
46 & 53536.4402 & 0.0006    & 120 & 53541.2451 & -0.0013  \\   
47 & 53536.5043 & -0.0003   & 121 & 53541.3107 & -0.0006  \\ 
48 & 53536.5701 & 0.0006    & 122 & 53541.3765 & 0.0002   \\ 
59 & 53537.2832 & -0.0009   & 123 & 53541.4421 & 0.0009    \\  
60 & 53537.3473 & -0.0017   & 124 & 53541.5029 & -0.0033   \\  
61 & 53537.4140 & 0.0000    & 125 & 53541.5735 & 0.0024   \\ 
62 & 53537.4781 & -0.0008   & 135 & 53542.2178 & -0.0029  \\ 
63 & 53537.5451 & 0.0012    & 136 & 53542.2819 & -0.0038   \\
73 & 53538.1905 & -0.0029   & 137 & 53542.3445 & -0.0061   \\
74 & 53538.2579 & -0.0005   & 138 & 53542.4118 & -0.0038   \\
75 & 53538.3219 & -0.0014   & 139 & 53542.4759 & -0.0046   \\
76 & 53538.3878 & -0.0005   & 140 & 53542.5414 & -0.0041   \\

\hline

\multicolumn{6}{@{}l@{}}{\hbox to 0pt{\parbox{180mm}{\footnotesize
\par\noindent
\footnotemark[$*$] Cycle count.
\par\noindent
\footnotemark[$\dagger$] Using equation (\ref{c}).
}\hss}}

\end{tabular}
\end{center}
\end{table*}

\section{DISCUSSION}

\subsection{\it Superhump Period}

During a superoutburst, the superhump period had been believed to decrease or stay constant in SU
UMa-type dwarf novae in general (e.g. \cite{war85suuma};
\cite{pat93vyaqr}), although only one unambiguous sign of
the increase of $P_{SH}$ was discovered during the superoutburst of OY Car
  \citep{krz85oycarsuper}.
 The decrease of the superhump period was explained by a shrinkage of the
 accretion disk (\cite{lub92SH}), or a inward-propagation of eccentric
 perturbations generated at the 3:1 resonance radius in the accretion disk (\cite{whi94SH}).

However, since \citet{sem97swuma} reported that SW UMa had shown an
increasing trend of
$P_{SH}$ during the 1996 superoutburst, 
researchers have revealed that some SU UMa-type dwarf novae exhibit the
increase of $P_{SH}$.
Such systems are mainly WZ Sge-type dwarf
novae and SU UMa-type dwarf novae with short orbital periods. 
\citet{kat98super} proposed that the accretion disk of short orbital period systems can expand beyond
the 3:1 resonance radius during superoutbursts and the eccentricity wave
originated at the 3:1 resonance can propagate outward. The outward-propagation of
the eccentricity may lead to the phenomena of the increase of $P_{SH}$ (see also \cite{kat04egcnc}). 
\citet{kat01hvvir} suggested that the increase of $P_{SH}$
appear to be related to a low mass ratio and/or a low mass transfer
rate.

Recently, \citet{osa03DNoutburst} classified superoutbursts of SU
UMa-type dwarf novae based on the mass ratio ($q=M_{2}/M_{1}$) and disk radius, and showed that a
system with a large mass ratio ($\sim0.2$) can exhibit two types of superoutburst.
Following this idea, \citet{uem05tvcrv} explained the behavior of the
superoutbursts of TV Crv\footnote{Kato et al. (in prep.) reanalyzed the
data of the superoutbursts of TV Crv and suggest that the interpretation
of the superhump period by \citet{uem05tvcrv} could not be confirmed.}. 
They proposed that the changing of $P_{SH}$
($P_{dot}$) depends on whether a
superoutburst has a precursor or not.
Their interpretation is indicated below: 
if a large amount of matter is accumulated beyond
the 3:1 resonance radius at the outburst maximum, the dammed matter prevents the disk from rapid
cooling, leading to a superoutburst without a precursor. 
In this case, the eccentricity originated
at the 3:1 resonance radius can propagate outward, and $P_{SH}$ can increase during
the superoutburst. 
Further, the discovery of the infrared
excess observed in the late stage of WZ Sge-type superoutburst also
supports their interpretation for the changing of $P_{SH}$ (\cite{uem08j1021}). 

In ASAS 1600, the $P_{SH}$ changed more
complexly than had been documented.
Figure \ref{o-c} represents an $O-C$ diagram for
the timings of the superhump maxima listed in table \ref{maxtime}. 
As we mentioned in section \ref{sh}, in the middle part (around
$30<E<100$), the $O-C$ diagram demonstrates an
increasing trend of $P_{SH}$, and the $P_{SH}$ derivative between
$28<E<109$ is 
$P_{dot}=\dot{P}_{SH}/P_{SH}=11.5(9) \times 10^{-5}$ days cycle$^{-1}$.
However, in the early (around $E=30$) and the last part (around
$E=100$), the $O-C$ diagram shows a decreasing trend of $P_{SH}$.
Some well observed SU UMa stars, e.g. AL Com
(\cite{how96alcom}, \cite{pat96alcom}), HV
Vir (\cite{ish03hvvir}) and TT
Boo (\cite{ole04ttboo}), showed an $O-C$ diagram with
the same trend as that of ASAS 1600. These systems seem to follow the
general tendency pointed out
by \citet{ole03ksuma}. They investigated the $O-C$ diagrams of some SU
UMa-type dwarf novae, and proposed that almost all SU
UMa-type dwarf novae show the decrease of $P_{SH}$ both at the beginning
and at the end of the superoutburst, but the increasing trend at the
middle phase.

We listed SU UMa-type dwarf novae which show an increase of
$P_{SH}$ so far (table \ref{positivepdot}). Because we judged from the shape of
the $O-C$ diagram, they include systems which have not been regarded as 
such. We reexamined them, and found that most
of the well
observed SU UMa-type dwarf novae with an increase of $P_{SH}$ also show
a decrease of $P_{SH}$ at the beginning or end stage, or both. This also
supports the claim by \citet{ole03ksuma}.

\subsection{\it Superhump Evolution}
\label{shevolution}

ASAS 1600 showed a regrowth of the
ordinary superhumps in the amplitude like
well investigated systems with an increase of $P_{SH}$ as described in
the previous section. On HJD
2453535, the superhump amplitude became
$\sim0.24$ mag at the first maximum. After that, it
gradually diminished, and reached $\sim0.13$ mag on HJD
2453539. However, the superhump amplitude regrew up to the second
maximum of  $\sim0.17$ mag on HJD
2453540, thereafter, it gradually became smaller again.

Additionally, from figure \ref{shamp} and \ref{o-c},
we found that the superhump period decreased when the superhump
amplitude reached to the first maximum, successively gradually
increased until the second maximum of the amplitude, and finally
decreased again during the 2005 superoutburst of ASAS 1600.
The correlation between the amplitude and the period of the superhumps can
be found in most SU UMa-type dwarf novae which show an increase of
$P_{SH}$ (see the systems listed in table \ref{positivepdot} with both
$*$ and $\dagger$ or
$\ddagger$). Therefore, we suppose that the regrowth of the superhump
amplitude have a close relation to the increase of $P_{SH}$, and 
most SU UMa-type dwarf novae with a
superhump regrowth may follow the same evolution of the ordinary
superhump as ASAS 1600. The statement about the
superhump period by \citet{ole03ksuma} seems to be applicable to such systems.

RZ LMi, however, showed only a gradual increase
of the superhump period during the whole course of the 2004 April
superoutburst, and this star does not seem to follow our scenario 
(\cite{ole08rzlmi})
\footnote{According to Rutkowski et al. (submitted to A\&A), one of ER
UMa stars, DI UMa also showed only an increase of the
superhump period.}. 
RZ LMi is an ER UMa-type star, which is
the most active subgroup of SU UMa-type dwarf novae
(\cite{nog95rzlmi}). The superhumps
in ER UMa stars have been pointed out to be somewhat different
from typical ones in normal SU UMa stars (e.g. \cite{kat03erumaSH}),
and \citet{osa95rzlmi} suggested that the extreme superoutburst properties
of RZ LMi can not be simply explained in the frame of the
thermal-tidal instability model. We may need a different scheme
to rightly interpret the superhump evolution in ER UMa stars.

Anyway, the relation between the amplitude and the period of ordinary superhumps
can be a key point for an understanding of the evolution of accretion
disks during a superoutburst, and observations focused on them are
required for the future.

\subsection{\it Is ASAS 1600 a WZ Sge-type Dwarf Nova?}

The
double-peaked humps were discovered with a period very close to the orbital period in
the early stage of the 2005 superoutburst of ASAS 1600 (Paper I), and,
therefore, 
this object was identified as a promising candidate of WZ Sge-type dwarf
novae. In 2006 May, the next year of the superoutburst, normal outburst of ASAS 1600
 was observed by R. Stubbings with $V\sim14.2$ (vsnet-outburst 6880), which seems to contradict the property (3) of WZ
 Sge-type dwarf novae. However,
AL Com, one of the established WZ Sge-type dwarf novae, also have undergone
some normal outbursts (in 1961, 1965, 1974, 1975, and possibly 1976
(\cite{how88faintCV1}, \cite{nog97alcom})), hence, the WZ Sge-type
identification of ASAS 1600 can not judged to be wrong only by this point.

\citet{kat08wzsgelateSH} proposed that, in
WZ Sge-type dwarf novae with
an extremely small mass ratio ($q=M_{2}/M_{1}$), the 2:1 resonance can be
strong enough to suppress the outward-propagation of the eccentricity originated at
the 3:1 resonance and, therefore, the $P_{SH}$ is almost constant. On the
other hand, in systems with a similar or
slightly larger $q$, such as RZ Leo
(\cite{ish01rzleo}), BC UMa (\cite{mae07bcuma})
and some SU UMa-type stars, the
outward-propagation of the eccentricity is not restricted and, therefore, the $P_{SH}$ can
be increase.
In Paper I, using an empirical relation (\cite{pat98egcnc}), authors
estimated the mass ratio of ASAS
1600 to be $q=0.109(4)$ which is slightly large, compared with that of the
typical WZ Sge-type dwarf novae. In section \ref{sh}, we determined the
$P_{dot}$ of ASAS 1600 to be $11.5(9) \times 10^{-5}$ days cycle$^{-1}$. Based on the idea by \citet{kat08wzsgelateSH}, these results indicate that ASAS
1600 is a system with a weak 2:1 resonance. 
ASAS 1600 does not seem to be a typical WZ Sge-type dwarf nova, but, at least, a very close to WZ Sge-type dwarf nova.

\begin{table*}
\begin{center}
\caption{SU UMa-type dwarf novae with the $P_{SH}$ increasing phase.}
\label{positivepdot}
\begin{tabular}{llcl}  
\hline \hline
 Object & $P_{SH}$ (days) & $P_{dot}^{\S}$ & Reference \\ 
\hline 
 
 V485 Cen $^{*}$ & 0.04216 & 28(3) & \citet{ole97v485cen} \\
 EI Psc & 0.04637 & 12(2) & \citet{uem02j2329} \\
 VS 0329+1250 & 0.053394 & 2.1(0.8) & \citet{sha07vs0329} \\
 V844 Her (2002) & 0.05584 & 4.4(1.2) & \citet{oiz07v844her} \\
 V844 Her (2006)$^{*\ddagger}$ & 0.055883 & 10.9(1.0) & \citet{oiz07v844her} \\
 AL Com (1995)$^{*\dagger\ddagger}$ & 0.0572 & 1.9(0.5)$^{\|}$ & \citet{how96alcom}, \citet{pat96alcom} \\
 ASAS J002511+1217.2 $^{*\ddagger}$ & 0.05687 & 8.7(0.4)$^{\|}$ & \citet{tem06asas0025} \\
 WZ Sge (2001)$^{*}$ & 0.05736 & 0.2(0.3)$^{\|}$ & \citet{pat02wzsge} \\
 SW UMa (1996)$^{*}$& 0.05818 & 8.9(1.0) & \citet{sem97swuma},
 \citet{nog98swuma} \\
 SW UMa (2000)$^{*\dagger\ddagger}$ & 0.058096 & 6.9(0.4) & Soejima et al. (in prep.) \\
 SW UMa (2002)$^{*\ddagger}$ & 0.058261 & 9.1(0.7) & Soejima et al. (in
 prep.) \\
 SW UMa (2006)$^{\dagger}$ & 0.058063 & 7.9(1.4) & Soejima et al. (in prep.) \\
 HV Vir (1992) & 0.05820 & 5.7(0.6) & \citet{kat01hvvir} \\
 HV Vir (2002)$^{\dagger\ddagger}$ & 0.05826 & 7.8(7) & \citet{ish03hvvir} \\
 RZ LMi (2004 Apr) & 0.05944 & 7.6(1.9) & \citet{ole08rzlmi} \\
 RZ LMi (2005 Apr) & 0.05940 & 4.5(2.5) & \citet{ole08rzlmi} \\
 WX Cet (1998)$^{*\ddagger}$ & 0.05949 & 8.5(1.0) & \citet{kat01wxcet} \\
 WX Cet (2001)$^{*}$ & 0.059563 & 16 & \citet{ste07wxcet} \\
 WX Cet (2004)$^{\ddagger}$ & 0.059585 & 16 & \citet{ste07wxcet} \\
 FL TrA $^{\dagger}$ & 0.059897 & 8.4(5.0) & \citet{ima08fltractcv0549} \\
 EG Cnc & 0.06043 & 1.7(1) & \citet{kat04egcnc} \\
 V1028 Cyg $^{*\dagger\ddagger}$ & 0.06154 & 8.7(0.9) & \citet{bab00v1028cyg} \\
 BC UMa $^{\ddagger}$ & 0.06258 & 3.2(0.8) & \citet{mae07bcuma} \\
 GO Com $^{*\ddagger}$ & 0.06306 & 18(3) & \citet{ima05gocom} \\
 V1159 Ori $^{*\dagger\ddagger}$ & 0.0641 & 4.2(1.1)$^{\|}$ & \citet{pat95v1159ori} \\
 OY Car (1980) & 0.064631 & 8.9(1.6)$^{\|}$ & \citet{krz85oycarsuper} \\
 ASAS 1600 $^{*\dagger\ddagger}$ & 0.064927 & 11.5(9) & this work \\
 TV Crv (2001) & 0.06503 & 8.0(0.7) & \citet{uem05tvcrv} \\
 KS UMa (2003)$^{*\ddagger}$ & 0.07009 & 21(12) & \citet{ole03ksuma} \\
 RZ Sge (1996)$^{\ddagger}$ & 0.07039 & 0.6(5.1)$^{\|}$ & \citet{sem97rzsge} \\
 VW Crb $^{*}$ & 0.07287 & 9.3(0.9) & \citet{nog04vwcrb} \\
 TT Boo (2004)$^{*\dagger\ddagger}$ & 0.07796 & 12.3(4.8) & \citet{ole04ttboo} \\
 RZ Leo $^{*\ddagger}$ & 0.07853 & 5.9(1.0) & \citet{ish01rzleo} \\

\hline

\multicolumn{4}{@{}l@{}}{\hbox to 0pt{\parbox{180mm}{\footnotesize
\par\noindent
\footnotemark[$*$] Dwarf novae with a regrowth of ordinary superhumps.
\par\noindent
\footnotemark[$\dagger$] Dwarf novae which show a decrease of $P_{SH}$ at the beginning stage.
\par\noindent
\footnotemark[$\ddagger$] Dwarf novae which show a decrease of $P_{SH}$ at the end stage.
\par\noindent
\footnotemark[$\S$] The unit is $10^{-5}$ days cycle$^{-1}$.
\par\noindent
\footnotemark[$\|$] Reference to Kato et al. (in prep.).
}\hss}}

\end{tabular}
\end{center}
\end{table*}

\section{CONCLUSION}

We photometrically studied the 2005 superoutburst of a dwarf nova ASAS
1600. Our conclusion is summarized below:

\begin{enumerate}

\item  ASAS 1600 showed a regrowth of the ordinary superhumps in the
       amplitude during the 2005 superoutburst, which is frequently
       found in SU UMa-type dwarf novae with an increase of $P_{SH}$.

\item  During the 2005 superoutburst, the superhump period decreased when
       the superhump amplitude reached to the first maximum, successively
       gradually increased until the second maximum of the amplitude,
       and finally decreased again. This course of the superhump
       evolution seems to be common in SU UMa-type dwarf novae with a
       superhump regrowth.

\item  ASAS 1600 does not seem to be a typical WZ Sge-type dwarf nova,
       but, at least, a very close to typical WZ Sge-type dwarf nova.

\end{enumerate}

\bigskip 
 
We acknowledge with thanks the variable star
observations from the AAVSO and VSNET International Database
contributed by observers worldwide and used in this research. 
Thanks are also due to the anonymous referee for useful comments.
This work was supported by the Grant-in-Aid for the Global COE Program
"The Next Generation of Physics, Spun from 
Universality and Emergence"
from the Ministry of Education, Culture, Sports, Science and Technology
(MEXT) of Japan.

\end{document}